\documentclass[prb,floatfix,showpacs]{revtex4}

\usepackage{graphicx}
\usepackage{dcolumn}
\usepackage{bm}

\begin{document}

\newcommand{\base}{}
\newcommand{\nonb}{\nonumber}
\newcommand{\RA}{\rangle}
\newcommand{\LA}{\langle}
\newcommand{\LL}{\langle \langle}
\newcommand{\RR}{\rangle \rangle}
\newcommand{\HG}{\hat{G}}
\newcommand{\HK}{\hat{K}}
\newcommand{\Ht}{\hat{t}}
\newcommand{\HA}{\hat{A}}
\newcommand{\HI}{\hat{I}}
\newcommand{\HX}{\hat{X}}
\newcommand{\HY}{\hat{Y}}
\newcommand{\Hg}{\hat{g}}
\newcommand{\Hrho}{\hat{\rho}}
\newcommand{\HSigma}{\hat{\Sigma}}
\newcommand{\CHI}{\check{I}}
\newcommand{\CHG}{\check{G}}
\newcommand{\CHK}{\check{K}}
\newcommand{\CHg}{\check{g}}
\newcommand{\DA}{{\cal D}^A}
\newcommand{\DR}{{\cal D}^R}
\newcommand{\CG}{{\cal G}}
\newcommand{\D}{{\cal D}}
\newcommand{\W}{{{\cal W}_{n,m}(N_a,N_b)}}
\newcommand{\vk}{{\vec{k}}}
\newcommand{\vx}{{\vec{x}}}
\newcommand{\be}{\begin{equation}}
\newcommand{\ee}{\end{equation}}

\newenvironment{tab}[1]
{\begin{tabular}{|#1|}\hline}
{\hline\end{tabular}}

\newcommand{\fig}[2]{\epsfxsize=#1\epsfbox{#2}} \reversemarginpar 
\bibliographystyle{prsty}

\title{Crossed Andreev reflection 
at ferromagnetic domain walls}
\author{
R. M\'elin\thanks{melin@grenoble.cnrs.fr}}
\affiliation{
Centre de Recherches sur les Tr\`es basses
temp\'eratures (CRTBT)\thanks{U.P.R. 5001 du CNRS,
Laboratoire conventionn\'e avec l'Universit\'e Joseph Fourier}\\
BP 166, 38042 Grenoble Cedex 9, France}

\author{S. Peysson}
\affiliation{Instituut voor Theoretische Fysica,
Universiteit van Amsterdam, \\
Valckenierstraat 65, 1018XE Amsterdam, The Netherlands}

\begin{abstract}
We investigate several factors controlling the physics of
hybrid structures involving ferromagnetic domain walls (DWs)
and superconducting (S) metals.
We discuss the role of non collinear magnetizations
in S/DW junctions in a spin $\otimes$
Nambu $\otimes$ Keldysh formalism. We discuss 
transport in S/DW/N and
S/DW/S junctions in the presence of inelastic scattering in the
domain wall. In this case transport properties are similar for
the S/DW/S and S/DW/N junctions and are controlled by sequential
tunneling of spatially separated Cooper pairs across
the domain wall.
In the absence of inelastic scattering we 
find that a Josephson current
circulates only if the size of the ferromagnetic region is smaller
than the elastic mean free path meaning that the Josephson effect
associated to crossed Andreev reflection cannot be observed under
usual experimental conditions. Nevertheless a finite dc current
can circulate across the S/DW/S junction due to 
crossed Andreev reflection associated to sequential
tunneling.
\end{abstract}


\maketitle

\section{Introduction}

A simple way of obtaining correlated pairs of electrons 
in solid state devices is to extract Cooper pairs from
a BCS superconductor. Devices based on
this principle have focussed an
important interest recently. For instance entangled
pairs of electrons can be manipulated in double
dot experiments~\cite{Loss}.
Other devices involving
a larger number of quantum dots have been proposed
recently as a quantum teleportation
experiment~\cite{quantum-tele}.
Devices involving several ferromagnetic electrodes
connected to a superconductor have been investigated
recently~\cite{Feinberg,Falci,Melin,Melin-Feinberg}.
Noise correlations can also provide useful information
about quantum entanglement~\cite{Martin}.

Many phenomena are involved in the proximity effect
at ferromagnet~/ superconductor (F/S) interfaces.
For instance
it is well established that the pair amplitude induced in
a ferromagnetic metal oscillates in space. An interesting
consequence is the possibility of fabricating S/F/S
$\pi$-junctions in which the Josephson relation 
is $I=I_c \sin{(\varphi+
\pi)}$~\cite{Fulde-Ferrel,Larkin,Clogston,Demler,Buzdin1,Ryazanov,Kontos,Zareyan}.
In F/S/F trilayers the superconducting transition temperature
is larger in the antiferromagnetic alignment of the
ferromagnetic electrodes~\cite{deGennes,Baladie} because a finite
exchange field is induced in the superconductor in the
ferromagnetic alignment. On the other hand there exist
``non local'' superconducting correlations coupling the two
ferromagnetic electrodes that favor $\Delta_{\rm F}
>\Delta_{\rm AF}$ (the zero-temperature superconducting
order parameter
is larger in the ferromagnetic alignment)~\cite{Apinyan,Jirari}.
It is also well known that the superconducting transition
temperature of F/S multilayers oscillates as the thickness
of the ferromagnetic layers is
increased~\cite{Buzdin2,exp1,exp2,exp3,exp4,exp5}.
Several recent works have investigated
new phenomena taking place in diffusive F/S
heterostructures~\cite{Lawrence,Vasko,Giroud0,Petrashov2,Filip,Giroud}.
Other recent works were devoted to understand the interplay
between Andreev reflection and spin polarization
at a single F/S interface~\cite{deJong,Soulen,Upa}.

In a recent article M. Giroud {\sl et al.}
have proposed on the basis of experiments
that the proximity effect at F/S 
interfaces could be strongly
modified by the presence of Cooper pair-like states
propagating along domain walls (DWs)~\cite{Giroud}.
These Cooper pair-like states correspond to
pair states in which the spin-up and
spin-down electrons propagate in a neighboring spin-up
and spin-down magnetic domain. This 
proximity effect is not strictly speaking equivalent
to the proximity effect a N/S interfaces. The reason
is that the pair correlations induced in the N side
of a N/S interface have
entangled orbital and spin degrees of freedom~\cite{Martin}.
By contrast for half-metal ferromagnets
the wave function associated to the 
propagation of superconducting correlations along
domain walls is given by the product state
$|e,\alpha ,\uparrow \rangle \otimes
|e ,\beta, \downarrow \rangle$, where $\alpha$
and $\beta$ represent two points in neighboring
magnetic domains. Another difference between a N/S
interface and a multiterminal hybrid structure is that
the incoming electron and the Andreev reflected
hole propagate in different electrodes
in multiterminal structures. As a consequence the
Andreev reflected hole cannot follow the same trajectory
as the incoming electron. This has important consequences
regarding disorder averaging.

The purpose of our article is to investigate
theoretically 
the mechanisms by which the Cooper pair-like state
$|e,\alpha ,\uparrow \rangle \otimes
|e ,\beta, \downarrow \rangle$ can propagate
along a ferromagnetic domain wall and to investigate
several new situations that may be the object
of experiments in the future. 
In section~\ref{sec:SDW} we discuss the perturbative
transport formula of a S/DW junction in which the domain
wall consists of many independent channels in parallel
having a rotating magnetization. To discuss this model
we use the spin $\otimes$ Nambu
$\otimes$ Keldysh formalism described in
section~\ref{sec:prelim}. For the sake of obtaining
analytical results we restrict the discussion to the
transport formula obtained within lowest order
perturbation theory. 

If propagation in the ferromagnet is
phase coherent then the pair state $|e,\alpha ,\uparrow
\rangle \otimes
|e ,\beta, \downarrow \rangle$ injected at one end of
the domain wall can propagate to the other end.
On the other hand if the phase coherence length $l_\phi$
is small compared to the size of the ferromagnetic region
then inelastic scattering
processes are strong and there are just a spin-up
and a spin-down electron propagating independently in the
spin-up and spin-down magnetic domains.
There is no Josephson current but there exists crossed
Andreev reflection taking place locally at each F/S interface,
so that
the conductance is larger in the presence of the domain wall.

In section~\ref{sec:out} we discuss the S/DW/N and
S/DW/S junctions in a regime where transport properties are
dominated by inelastic scattering in the domain wall.
The domain wall is represented
by two channels in parallel, with an opposite magnetization.
This schematic model of domain wall is expected to capture the
essential physics, and can be a useful comparison for more
realistic studies involving numerical simulations that we plan
to carry out in the future. We show that within lowest
order perturbation
the transport properties are governed by
processes taking place locally at each interface once the
summation over the different conduction channels has been
carried out.
The chemical potentials in the domain wall
are determined by evaluating the current circulating through
each interface and imposing current conservation.

In section~\ref{sec:SDWS-Jo} we consider the other situation
where inelastic scattering within the domain wall can be neglected.
In this situation a finite average
Josephson current can circulate between
the two superconductors of the S/DW/S junction only if the
size of the ferromagnetic region is smaller than the elastic
mean free path. This condition is not realized with usual ferromagnets
and we come to the conclusion that there is no Josephson current
under usual experimental conditions.
Final remarks are given in section~\ref{sec:conclu}.

To end-up the introductory section we note that the theory of
inhomogeneous ferromagnets with non colinear
magnetizations in contact with a superconductor was already
elaborated in Refs.~\cite{Bergeret,Kadi} in connection with
the long-range proximity effect associated to the triplet
component of the superconducting condensate. In our
article the emphasis is put on other aspects of this
problem (the transport of spatially separated Cooper pairs).
Both effects may play a relevant role in experiments.
Finally a recent preprint~\cite{Cht} appeared in which the
conductance of a S/DW junction was calculated independtly from our work.

\begin{figure}
\includegraphics [width=.3 \linewidth]{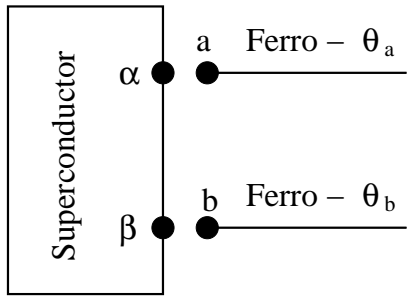}
\caption{The device involving crossed Andreev 
reflection and elastic cotunneling with
non collinear magnetizations. Electrode ending at site ``a''
is a ferromagnet with a magnetization pointing in
the direction $\theta_a$.
Electrode ending at site ``b'' is a 
ferromagnet with a magnetization pointing
in the direction $\theta_b$.
}
\label{fig:schema1}
\end{figure}

\section{Preliminaries}
\label{sec:prelim}
\label{sec:forma}

\subsection{Spin $\otimes$ Nambu $\otimes$ Keldysh formalism}

The direction of the magnetization is rotating in a
ferromagnetic domain wall. 
To describe superconducting correlations in
the presence of
non collinear magnetizations we use a
spin $\otimes$ Nambu $\otimes$ Keldysh
formalism~\cite{Feigelman,circuit,Zeng}.
The advanced Green's function is a $4 \times 4$ matrix:
\begin{equation}
\hat{G}^A_{i,j}(t,t') = -i \theta(t-t')
\left[ \begin{array}{cccc}
\langle \left\{ c_{j,\uparrow}^+(t') , c_{i,\uparrow}(t) \right\} \rangle &
\langle \left\{ c_{j,\downarrow}(t') , c_{i,\uparrow}(t) \right\} \rangle &
\langle \left\{ c_{j,\downarrow}^+(t') , c_{i,\uparrow}(t) \right\} \rangle 
&
\langle \left\{ c_{j,\uparrow}(t') , c_{i,\uparrow}(t) \right\} \rangle \\
\langle \left\{ c_{j,\uparrow}^+(t') , c_{i,\downarrow}^+(t) \right\} 
\rangle &
\langle \left\{ c_{j,\downarrow}(t') , c_{i,\downarrow}^+(t) \right\} 
\rangle &
\langle \left\{ c_{j,\downarrow}^+(t') , c_{i,\downarrow}^+(t) \right\} 
\rangle &
\langle \left\{ c_{j,\uparrow}(t') , c_{i,\downarrow}^+(t) \right\} \rangle 
\\
\langle \left\{ c_{j,\uparrow}^+(t') , c_{i,\downarrow}(t) \right\} \rangle 
&
\langle \left\{ c_{j,\downarrow}(t') , c_{i,\downarrow}(t) \right\} \rangle 
&
\langle \left\{ c_{j,\downarrow}^+(t') , c_{i,\downarrow}(t) \right\} 
\rangle &
\langle \left\{ c_{j,\uparrow}(t') , c_{i,\downarrow}(t) \right\} \rangle \\
\langle \left\{ c_{j,\uparrow}^+(t') , c_{i,\uparrow}^+(t) \right\} \rangle 
&
\langle \left\{ c_{j,\downarrow}(t') , c_{i,\uparrow}^+(t) \right\} \rangle 
&
\langle \left\{ c_{j,\downarrow}^+(t') , c_{i,\uparrow}^+ \right\} \rangle &
\langle \left\{ c_{j,\uparrow}(t') , c_{i,\uparrow}^+(t) \right\} \rangle \\
\end{array} \right]
.
\end{equation}
The Dyson equation relates the Green's functions of the
connected system to the Green's functions of the
disconnected system.
In a compact notation
the Dyson equation takes the form $\HG = \Hg
+ \Hg \otimes \hat{\Sigma} \otimes \HG$, where the
symbol $\otimes$ includes a summation over the sites
of the network and a convolution over time variables.
Since we consider stationary transport the convolution
over time variables
becomes a simple product after a Fourier transform
is carried out.
The Dyson equation for the Keldysh Green's function
$\hat{G}^{+,-}$ is given by
\begin{equation}
\label{eq:Dyson-Keldysh}
\HG^{+,-} = \left[ \hat{I} +
\hat{G}^R \otimes \hat{\Sigma} \right] \otimes
\hat{g}^{+,-} \otimes
\left[ \hat{I} + \hat{\Sigma}
\otimes \hat{G}^A \right]
,
\end{equation}
where the self-energy $\hat{\Sigma}$ contains all the
couplings present in the tunnel Hamiltonian.
The tunnel Hamiltonian corresponding to
Fig.~\ref{fig:schema1} takes the form
\begin{equation}
{\cal W} = \sum_\sigma \left[
t_{a,\alpha} c_{a,\sigma}^+ c_{\alpha,\sigma}
+ t_{\alpha,a} c_{\alpha,\sigma}^+ c_{a,\sigma}
+t_{b,\beta} c_{b,\sigma}^+ c_{\beta,\sigma}
+t_{\beta,b} c_{\beta,\sigma}^+ c_{b,\sigma}
\right]
.
\end{equation}
The current through the link $a$~--~$\alpha$ is given by
\begin{equation}
\label{eq:transport-formula}
I_{a,\alpha}=\frac{e}{2h} \int 
\mbox{Tr} \left\{ \hat{\sigma}_z \left[
\hat{t}_{a,\alpha} \hat{G}^{+,-}_{\alpha,a}
-\hat{t}_{\alpha,a} \hat{G}^{+,-}_{a,\alpha}
\right] \right\} d\omega
,
\end{equation}
where the matrix
$\hat{\sigma}_z$ is given by
\begin{equation}
\hat{\sigma}_z = 
\left[ \begin{array}{cccc}
1 & 0 & 0 & 0 \\
0 & -1 & 0 & 0 \\
0 & 0 & 1 & 0 \\
0 & 0 & 0 & -1
\end{array} \right]
\end{equation}
and the Nambu representation of the 
hopping matrix elements is given by
$\hat{t}_{a,\alpha} = t_{a,\alpha} \hat{\sigma}_z$,
$\hat{t}_{\alpha,a} = t_{\alpha,a} \hat{\sigma}_z$,
$\hat{t}_{b,\beta} = t_{b,\beta} \hat{\sigma}_z$,
$\hat{t}_{\beta,b} = t_{\beta,b} \hat{\sigma}_z$.

\subsection{Green's function of a ferromagnetic metal}

Now we give the expressions of the Green's functions of a 
ferromagnetic metal. We first suppose that the spin quantization
axis is parallel to the direction of the magnetization. 
The Green's function takes the form
\begin{equation}
\Hg(R,\omega) = \left[ \begin{array}{cccc}
g_{1,1}(R,\omega) & 0 & 0 & 0 \\
0 & g_{2,2}(R,\omega) & 0 & 0 \\
0 & 0 & g_{3,3}(R,\omega) & 0 \\
0 & 0 & 0 & g_{4,4}(R,\omega)
\end{array} \right]
.
\end{equation}
The four diagonal elements are given by
\begin{eqnarray}
\label{eq:g11}
g_{1,1}(R,\omega) &=&- \frac{m_\uparrow a_0^2}{\hbar^2}
\frac{a_0}{2\pi R} \exp{\left\{-i \left( k_F^\uparrow
+\frac{\omega}{v_F^\uparrow} \right) R \right\} }
\exp{\left\{-\left(\frac{R}{l_\phi}\right) \right\}}
\\
\label{eq:g22}
g_{2,2}(R,\omega) &=& \frac{m_\downarrow a_0^2}{\hbar^2}
\frac{a_0}{2\pi R} \exp{\left\{i \left( k_F^\downarrow
-\frac{\omega}{v_F^\downarrow} \right) R \right\} }
\exp{\left\{-\left(\frac{R}{l_\phi} \right)\right\}}
\\
\label{eq:g33}
g_{3,3}(R,\omega) &=& - \frac{m_\downarrow a_0^2}{\hbar^2}
\frac{a_0}{2\pi R} \exp{\left\{-i \left( k_F^\downarrow
+\frac{\omega}{v_F^\downarrow} \right) R \right\} }
\exp{\left\{-\left(\frac{R}{l_\phi} \right)\right\}}
\\
\label{eq:g44}
g_{4,4}(R,\omega) &=&  \frac{m_\uparrow a_0^2}{\hbar^2}
\frac{a_0}{2\pi R} \exp{\left\{i \left( k_F^\uparrow
-\frac{\omega}{v_F^\uparrow} \right) R \right\} }
\exp{\left\{-\left(\frac{R}{l_\phi} \right)\right\}}
,
\end{eqnarray}
where we have introduced a Fermi wave vector mismatch as well
as a mismatch between the spin-up and spin-down Fermi velocities.
The parameter $a_0$ is equal to the distance between neighboring
sites on the cubic lattice.
For generality we introduced a different mass for the
spin-up and spin-down electrons, meaning that the
spin-up density of states is different from
the spin-down density of states.
The local propagators are defined by
\begin{eqnarray}
\label{eq:g-loc}
g_{1,1}^{\rm loc} &=& g_{4,4}^{\rm loc} =
i \frac{a_0 k_F^\uparrow}{2 \pi}
\frac{m_\uparrow a_0^2}{\hbar^2}=i\pi \rho_F
\left(\frac{1+P}{2}\right)\\
g_{2,2}^{\rm loc} &=& g_{3,3}^{\rm loc} =
i \frac{a_0 k_F^\downarrow}{2 \pi}
\frac{m_\downarrow a_0^2}{\hbar^2}=i\pi \rho_F
\left(\frac{1-P}{2}\right)
.
\end{eqnarray}
We also introduced phenomenologically
in (\ref{eq:g11})~-- (\ref{eq:g44}) an exponential decay
of the correlations due to the presence of a finite
coherence length $l_\phi$ in the ferromagnet.
$l_\phi$ is usually smaller than the
dimension of the ferromagnetic metal.
In this case ferromagnetism can be treated 
semi-classically like in the theoretical description
of the giant
magnetoresistance~\cite{Valet-Fert,Gijs-Bauer,Melin-Denaro}.
However Aharonov-Bohm oscillations in a ferromagnetic
nanoring have been reported recently~\cite{Kasai}.
The inner diameter of the Fe-Ni nanoring in Ref.~\cite{Kasai}
is $420 \AA$ and the outer diameter is $500 \AA$.

We will use in section~\ref{sec:SDW}
the expression of the local Green's functions of 
a ferromagnetic metal with the quantization axis not 
parallel to the magnetization.
We suppose that the direction
of the exchange field is rotated by an angle $\theta$
around the $x$ axis. We do not incorporate
a rotation of angle $\varphi$ around the
$z$ axis since this rotation just introduces
simple phase factors.
The local Green's function
of the rotated ferromagnet takes the form
\begin{equation}
\hat{g}_{\rm loc} = i \pi \tilde{\rho}
\left[ \begin{array}{cccc}
1+P\cos{\theta} & 0 & -i P \sin{\theta} & 0 \\
0 & 1-P\cos{\theta} & 0 & -i P \sin{\theta} \\
i P \sin{\theta} & 0 & 1 - P \cos{\theta} & 0 \\
0 & i P \sin{\theta} & 0 & 1 + P \cos{\theta}
\end{array} \right]
,
\end{equation}
where $\tilde{\rho}=(\rho_\uparrow
+\rho_\downarrow)/2$ is the average density of
states at the Fermi level and
$P = (\rho_\uparrow - \rho_\downarrow) /
(\rho_\uparrow + \rho_\downarrow)$ is the
spin polarization at the Fermi level.

We will also use in section~\ref{sec:multi}
the expression of the full propagator
$\hat{g}(R,\omega)$ of a rotated ferromagnet.
The Green's function takes the form
\be
\hat{g}(R,\omega) = \left[
\begin{array}{cccc}
\tilde{g}_{1,1} & 0 & \tilde{g}_{1,3} & 0 \\
0 & \tilde{g}_{2,2} & 0 & \tilde{g}_{2,4} \\
\tilde{g}_{3,1} & 0 & \tilde{g}_{3,3} & 0 \\
0 & \tilde{g}_{4,2} & 0 & \tilde{g}_{4,4}
\end{array} \right]
,
\ee
where the diagonal elements are given by
\begin{eqnarray}
\tilde{g}_{1,1} &=& \frac{1}{2} (g_{1,1}
+g_{3,3}) + \frac{1}{2} \cos{\theta}
(g_{1,1}-g_{3,3})\\
\tilde{g}_{2,2} &=& \frac{1}{2}(g_{2,2}+g_{4,4})
+\frac{1}{2} \cos{\theta} (g_{2,2}-g_{4,4})\\
\tilde{g}_{3,3} &=& \frac{1}{2} (g_{3,3}+g_{1,1})
+\frac{1}{2} \cos{\theta} (g_{3,3}-g_{1,1})\\
\tilde{g}_{4,4} &=& \frac{1}{2}(g_{4,4}+g_{2,2})
+\frac{1}{2} \cos{\theta} (g_{4,4}-g_{2,2})
,
\end{eqnarray}
where $g_{1,1}$, $g_{2,2}$, $g_{3,3}$ and
$g_{4,4}$ are given by (\ref{eq:g11})-(\ref{eq:g44}).
The extra-diagonal elements are given by
\begin{eqnarray}
\tilde{g}_{1,3} &=& -\tilde{g}_{3,1}=\frac{i}{2}
\sin{\theta} (g_{3,3}-g_{1,1} )\\
\tilde{g}_{2,4} &=&
-\tilde{g}_{4,2} = \frac{i}{2} \sin{\theta}
(g_{2,2}-g_{4,4})
.
\end{eqnarray}

\subsection{$4 \times 4$ Green's functions of a superconductor}
\label{sec:4x4-supra}
The Green's function of a superconductor takes the form
\be
\Hg^{A,R}(R,\omega) = \left[ \begin{array}{cccc}
g(R,\omega) & f(R,\omega) & 0 & 0 \\
f(R,\omega) & g'(R,\omega) & 0 & 0 \\
0 & 0 & g(R,\omega) & -f(R,\omega) \\
0 & 0 & -f(R,\omega) & g'(R,\omega)  \end{array} \right]
.
\ee

The matrix elements of the Green's function are given by
\begin{eqnarray}
g(R,\omega) &=& 
\frac{m a_0^2}{\hbar^2}
\frac{a_0}{2\pi R} \exp{\left(-\frac{R}{\xi(\omega)}\right)}
\left\{  \sin{(k_F R)} \frac{-\omega}{\sqrt{\Delta^2-\omega^2}}
-\cos{(k_F R)} \right\}\\
g'(R,\omega) &=& 
\frac{m a_0^2}{\hbar^2}
\frac{a_0}{2\pi R} \exp{\left(-\frac{R}{\xi(\omega)}\right)}
\left\{  \sin{(k_F R)} \frac{-\omega}{\sqrt{\Delta^2-\omega^2}}
+\cos{(k_F R)} \right\}\\
\label{eq:f-supra}
f(R,\omega) &=& 
\frac{m a_0^2}{\hbar^2}
\frac{a_0}{2\pi R} \exp{\left(-\frac{R}{\xi(\omega)}\right)}
\sin{(k_F R)} \frac{\Delta}{\sqrt{\Delta^2-\omega^2}}
,
\end{eqnarray}
where we supposed that $\omega<\Delta$.
The coherence length is given by $\xi(\omega)=\hbar v_F/
\sqrt{\Delta^2-\omega^2}$.

\subsection{$4 \times 4$ Green's functions of a superconductor
in a uniform magnetic field}
\label{sec:g+}
A uniform magnetic field $h_S$ can penetrate in a
superconductor if the superconductor is in a thin film
geometry~\cite{Tedrow} and the magnetic field is
parallel to the direction of the
superconducting film. The effect of the magnetic field
is a Zeeman splitting of the spin-up and spin-down
quasiparticle bands. Let us suppose that the quantization
axis is parallel to the orientation of the magnetic field.
The $4 \times 4$ Green's function
takes the form
\be
g^{A,R}(R,\omega) = \left[ \begin{array}{cccc}
g_+(R,\omega) & f_+(R,\omega) & 0 & 0 \\
f_+(R,\omega) & g'_+(R,\omega) & 0 & 0 \\
0 & 0 & g_-(R,\omega) & -f_-(R,\omega) \\
0 & 0 & -f_-(R,\omega) & g'_-(R,\omega)
\end{array} \right]
,
\ee
with $g_+(R,\omega)=g(R,\omega+h_S)$,
$g'_+(R,\omega)=g'(R,\omega+h_S)$,
$f_+(R,\omega)=f(R,\omega+h_S)$,
$g_-(R,\omega)=g(R,\omega-h_S)$,
$g'_-(R,\omega)=g'(R,\omega-h_S)$,
$f_-(R,\omega)=f(R,\omega-h_S)$.

\section{Crossed Andreev reflection and elastic cotunneling
with non collinear magnetizations}
\label{sec:SDW}
\subsection{Transport formula}
In this section we evaluate the transport formula
corresponding to the device on Fig.~\ref{fig:schema1} in which
the magnetization of electrode ``a'' (``b'') makes an angle
$\theta_a$ ($\theta_b$) with the $z$-axis.
Using the formalism described in section~\ref{sec:forma}
we obtain the current per conduction channel
through electrode ``a''
to lowest order in $t_{a,\alpha}$ and $t_{b,\beta}$:
\begin{eqnarray}
I_{a,\alpha} &=& \frac{e}{h} \int d\omega 8 \pi^2 t_\alpha^4
\tilde{\rho}_a^2 (1-P_a^2) f_{\rm loc}^2(\omega)
\left[ n_F(\omega-e V_a) - n_F(\omega+e V_a) \right] \\\nonb
&+& \frac{e}{h} \int d\omega 4 \pi^2 t_\alpha^2
t_\beta^2 \tilde{\rho}_a \tilde{\rho}_b
\left[1+P_a P_b \cos{(\theta_a-\theta_b)}\right]
\LL g_{\alpha,\beta} g_{\beta,\alpha} \RR \\\nonb
&& \times \left[
n_F(\omega-e V_a) - n_F(\omega+e V_a)
-n_F(\omega-e V_b) +n_F(\omega + e V_b) \right]\\\nonb
&+& \frac{e}{h} \int d\omega 4 \pi^2 t_\alpha^2
t_\beta^2 \tilde{\rho}_a \tilde{\rho}_b
\left[1-P_a P_b \cos{(\theta_a-\theta_b)}\right]
\LL f_{\alpha,\beta} f_{\beta,\alpha} \RR \\\nonb
&& \times \left[
n_F(\omega-e V_a) - n_F(\omega+e V_a)
+n_F(\omega-e V_b) -n_F(\omega + e V_b) \right]
.
\end{eqnarray}
We have supposed that electrodes ``a'' and ``b''
are made of a large
number of independent conduction channels in parallel
so that we make an averaging over the
microscopic phases in the propagators.
Now if we consider that the same voltage is applied on
both electrodes 
the conductance is given by
local Andreev reflection and crossed Andreev reflection:
\begin{equation}
\label{eq:G-rotating}
G =  32 \pi^2 \frac{e^2}{h} t_\alpha^4 \tilde{\rho}_a^2
(1-P_a^2) f_{\rm loc}^2
+  32 \pi^2 \frac{e^2}{h} t_\alpha^2 t_\beta^2
\tilde{\rho}_a \tilde{\rho}_b 
\left[ 1-P_a P_b
\cos{(\theta_a-\theta_b)} \right]
\LL f_{\alpha,\beta} f_{\beta,\alpha} \RR
.
\end{equation}
\begin{figure}
\includegraphics [width=.3 \linewidth]{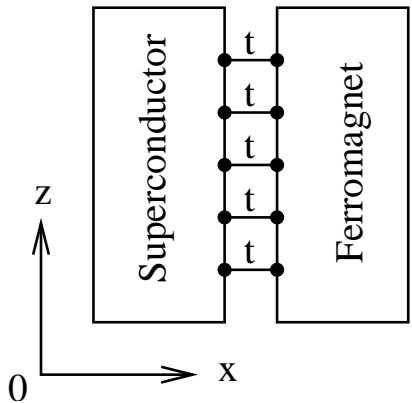}
\caption{The device involving a S/DW junction between a
superconductor and a ferromagnetic domain wall.
In the ferromagnet the local magnetization makes an angle
$\theta(z)$ with the $z$-axis. The $x$- and $z$-axis
are in the plane of the figure. The $y$ axis is
perpendicular to the plane of the figure.
}
\label{fig:schema2}
\end{figure}

\subsection{Conductance associated to a domain wall}
\label{sec:cond-DW}
Let us now
consider the situation on Fig.~\ref{fig:schema2}
representing a S/DW contact between a superconductor
and a magnetic domain wall. We suppose that the
ferromagnetic metal is made of a collection of independent
channels. The magnetization is rotating inside the domain
wall meaning that the angle $\theta$ is a function
of $z$: $\theta=\theta(z)$. We want to evaluate the
difference $G^{(DW)}-G^{(0)}$
between the conductances $G^{(DW)}$ in
the presence of the domain wall and $G^{(0)}$
in the absence of the domain wall.
To obtain the conductance 
we sum the contributions of the different channels
(see Fig.~\ref{fig:schema2}) and we obtain
\be
G^{(DW)}-G^{(0)} = 4 \frac{e^2}{h}
\frac{L_y}{a_0} t^4 \tilde{\rho}^2 P^2
\left( \frac{m a_0^2}{\hbar^2} \right)^2 
F(\xi_0,D)
,
\ee
with
\be
\label{eq:F}
F(\xi_0,D)=\int \frac{d (\Delta y)}{a_0}
\int \frac{d z_a}{a_0} \int \frac{d z_b}{a_0}
\frac{a_0^2}{ (\Delta y)^2+(z_a-z_b)^2}
\sin^2{\left(\frac{\theta(z_a)-\theta(z_b)}{2}\right)}
\exp{\left(-\frac{2
\sqrt{ (\Delta y)^2+(z_a-z_b)^2}}{\xi_0}\right)}
,
\ee
where $\xi_0=\hbar v_F / \Delta$ is the BCS coherence length
at zero energy and $D$ is the width of the domain wall.
$L_y$ is equal to the dimension of the junction in the $y$ direction and
we used the notation $\Delta y = y_a - y_b$.
To obtain (\ref{eq:F}) we
have supposed that the width of the
domain wall is much larger than the Fermi wave-length so
that we can average over the microscopic phase variables
in the propagator $f_{\alpha,\beta}$ (see Eq.~(\ref{eq:f-supra})).

\begin{figure}
\includegraphics [width=.4 \linewidth]{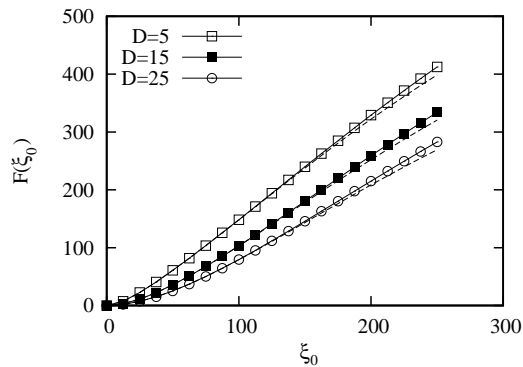}
\caption{Variation of $F(\xi_0,D)$ as a function of
$\xi_0$ for different values of $D$.
We used the domain wall profile given by
(\ref{eq:profile}) in a three-dimensional geometry.
The distance between neighboring channels is $a_0=1$.
We introduced a cut-off $-L_y/2 \le \Delta y \le L_y$,
$-L_z/2 \le z_a, z_b \le L_z/2$ in the expression
of $F(\xi_0,D)$ (see Eq.~(\ref{eq:F})). The solid
lines correspond to $L_y=L_z=500$ and the dashed line
(almost superimposed with the solid line)
correspond to $L_y=L_z=600$. 
}
\label{fig:DW}
\end{figure}

Crossed Andreev reflection cannot take place
between the channels
separated by a distance much smaller than the width
$D$ of the domain wall
because such channels have an almost parallel
magnetization. Crossed Andreev reflection cannot
take place either between channels separated by a distance
much larger than the superconducting coherence length because
of the exponential decay of the propagator $f_{\alpha,\beta}$.
As a consequence
the value of $G^{(DW)}-G^{(0)}$ is the largest if the
width of the domain wall is small compared to the BCS
coherence length. This is illustrated on Fig.~\ref{fig:DW}
where we have represented the variation of the conductance
as a function of $\xi_0$ for different values of $D$ and
for the domain wall profile given by
\begin{equation}
\label{eq:profile}
\theta(z) = \arctan{\left( \frac{z}{D} \right)}
.
\end{equation}

\subsection{Exchange field in the superconductor due to
the proximity effect}
\label{sec:precess}
Now we come back to a system in which two
ferromagnetic electrodes are connected to a superconductor.
An exchange field can be generated in the superconductor
because of the proximity effect. This was first observed
in Ref.~\cite{deGennes} in the case of insulating ferromagnets.
An exchange field in a superconductor is a pair breaking
perturbation.
As a consequence in the F/S/F trilayer with insulating
ferromagnets the order parameter is larger in the
antiferromagnetic alignment of the ferromagnetic
electrodes~\cite{deGennes}. This was well verified
in experiments with insulating
ferromagnets~\cite{Deutscher,Hauser}.
The same effect is present with metallic
ferromagnets~\cite{Baladie,Jirari} but in this case there
exists also pair correlations induced in the ferromagnetic
electrodes~\cite{Apinyan,Jirari} that can modify the value
of the self consistent order parameter. 

We suppose that the magnetizations in electrodes ``a''
and ``b'' make an angle $\theta_a$ and $\theta_b$ and
that an exchange field $h_S$ is induced in the superconductor.
Without loss of generality we suppose that the
direction of the exchange field in the superconductor
is parallel to the quantization axis.
In terms of the $g_+$, $f_+$ , $g_-$ and $f_-$ introduced
in section~\ref{sec:g+} the transport formula is found to be
\begin{eqnarray}
\label{eq:local-AR+}
I_{a,\alpha} &=& \frac{e}{h} \int d\omega
4 \pi^2 t_\alpha^4 \tilde{\rho}_a^2
\left[ \left( f_+^2 + f_-^2 \right)
\left(1-P_a^2 \cos^2{\theta_a} \right)
-2 f_+ f_- P_a^2 \sin^2{\theta_a} \right]
\left[ n_F(\omega-e V_a) - n_F(\omega + e V_a) \right]\\
\label{eq:CE+}
&+&\frac{e}{h} \int d\omega
2\pi^2 t_\alpha^2 t_\beta^2 \tilde{\rho}_a \tilde{\rho}_b
\left\{ \left[ \LL g^+_{\alpha,\beta} g^+_{\beta,\alpha} \RR
+\LL g^-_{\alpha,\beta} g^-_{\beta,\alpha} \RR \right]
\left[ 1+ P_a P_b \cos{\theta_a} \cos{\theta_b} \right]
\right.\\\nonb
&&+ \left. 2 \LL g^+_{\alpha,\beta} g^-_{\beta,\alpha} \RR
P_a P_b \sin{\theta_a} \sin{\theta_b} \right\}
\times \left[ n_F(\omega-e V_a)
-n_F(\omega+e V_a) -n_F(\omega-e V_b)
+n_F(\omega+e V_b) \right]\\
\label{eq:crossed-AR+}
&+& \frac{e}{h} \int d\omega
2\pi^2 t_\alpha^2 t_\beta^2 \tilde{\rho}_a \tilde{\rho}_b
\left\{ \left[ \LL f^+_{\alpha,\beta} f^+_{\beta,\alpha} \RR
+\LL f^-_{\alpha,\beta} f^-_{\beta,\alpha} \RR \right]
\left[ 1-P_a P_b \cos{\theta_a} \cos{\theta_b} \right] \right.\\\nonb
&&- \left. 2 \LL f^+_{\alpha,\beta} f^-_{\beta,\alpha} \RR
P_a P_b \sin{\theta_a} \sin{\theta_b} \right\}
\times \left[ n_F(\omega-e V_a)
-n_F(\omega+e V_a) + n_F(\omega-e V_b)
-n_F(\omega+e V_b) \right]
.
\end{eqnarray}
The term~(\ref{eq:local-AR+}) corresponds to
local Andreev reflection. The term~(\ref{eq:CE+})
corresponds to elastic cotunneling
and the term~(\ref{eq:crossed-AR+})
corresponds to crossed Andreev reflection.
The term $\LL g^+_{\alpha,\beta} g^+_{\beta,\alpha} \RR$
corresponds to a process in which a spin-up electron
travels from electrode $a$ to electrode $b$ and comes
back to electrode $a$ as a spin-up electron. The
term $\LL g^+_{\alpha,\beta} g^-_{\beta,\alpha} \RR$
corresponds to a process in which a spin-up electron
travels from electrode $a$ to electrode $b$, undergoes
a spin precession in electrode $b$ and comes back as
a spin-down electron traveling from electrode $b$
to electrode $a$.

Replacing the propagators involved in Eqs.(\ref{eq:local-AR+}),
(\ref{eq:CE+}) and (\ref{eq:crossed-AR+})
by their
expressions given in section~\ref{sec:4x4-supra}
leads to the transport formula
to lowest order in $h_S$ and $\omega$:
\begin{eqnarray}
I_{a,\alpha} &=& 8 \pi^2 t_\alpha^4
\left(\frac{m a_0^2}{\hbar^2} \right)^2
\left(\frac{a_0}{2\pi R_0} \right)^2
\left[ 1 + \frac{\omega^2 + h_S^2}{\Delta^2} \right]
\left[1 - P_a^2 \right] \left[ n_F(\omega-e V_a)
-n_F(\omega + e V_a) \right]\\
\label{eq:CAR-explicit}
&+& 2 \pi^2 t_\alpha^2 t_\beta^2
\tilde{\rho}_a \tilde{\rho}_b
\left( \frac{m a_0^2}{\hbar^2} \right)^2
\left( \frac{a_0}{2\pi R_{\alpha,\beta}} \right)^2
\exp{\left(-\frac{2 R_{\alpha,\beta}}{\xi(\omega)}\right)}
\left\{1 + \left[ 1 + \frac{\omega^2}{\Delta^2} \right]
P_a P_b \cos{(\theta_a-\theta_b)} \right.\\\nonb
&&+ \left. \frac{h_S^2}{\Delta^2} P_a P_b \cos{(\theta_a+\theta_b)}
\right\} 
\times
\left[ n_F(\omega-e V_a) - n_F(\omega+e V_a)
-n_F(\omega-e V_b) +n_F(\omega+e V_b) \right]\\
\label{eq:EC-explicit}
&+& 2 \pi^2 t_\alpha^2 t_\beta^2 \tilde{\rho}_a
\tilde{\rho}_b \left( \frac{m a_0^2}{\hbar^2} \right)
\left( \frac{a_0}{2\pi R_{\alpha,\beta}} \right)^2 
\exp{\left(-\frac{2 R_{\alpha,\beta}}{\xi(\omega)}\right)}
\left[ 1 + \frac{\omega^2+h_S^2}{\Delta^2} \right]
\left[ 1 - P_a P_b \cos{(\theta_a-\theta_b)} \right]\\\nonb
&& \times \left[ n_F(\omega-e V_a) - n_F(\omega+e V_a)
+n_F(\omega-e V_b) -n_F(\omega+e V_b) \right]
,
\end{eqnarray}
where $k_F R_0=\pi/2$ is the ultraviolet
cut-off used to define the local propagator involved
in local Andreev reflection. We see that the crossed
Andreev reflection term given by~(\ref{eq:CAR-explicit})
is not identical to the elastic cotunneling term
given by~(\ref{eq:EC-explicit}). This shows that
the symmetry between elastic cotunneling and crossed
Andreev reflection is broken by the exchange
field in the superconductor. This can be illustrated
by considering that electrode ``b'' is a normal metal:
$P_b=0$. The crossed conductance at zero voltage
is finite if the exchange field $h_S$ in the superconductor 
takes a finite value:
\be
G_{a,b}=\frac{\partial I_a}{\partial V_b}
=4 \pi^2 t_\alpha^2 t_\beta^2 \tilde{\rho}_a
\tilde{\rho}_b \left( \frac{m a_0^2}{\hbar^2} \right)^2
\left( \frac{a_0}{2 \pi R_{\alpha,\beta}} \right)^2
\exp{\left(-\frac{2 R_{\alpha,\beta}}{\xi_0} \right)}
\left( \frac{h_S}{\Delta} \right)^2
.
\ee
By comparison we have $G_{a,b}=0$ if $h_S=0$ because
of a cancellation between the crossed Andreev reflection
and elastic cotunneling conductances. We thus see that a
crossed Andreev reflection experiment with a ferromagnetic
and a normal metal electrode can give information about
the existence of an induced exchange field in the
superconductor.
We see also from Eqs.~(\ref{eq:local-AR+}),~(\ref{eq:CE+})
and~(\ref{eq:crossed-AR+})
that there is no precession of the
electron spin around the direction of the exchange field
in the superconductor. The absence of spin precession
in the superconducting case can be contrasted with the
metallic case (see Appendix~\ref{app:precession}).

\begin{figure}
\includegraphics [width=.4 \linewidth]{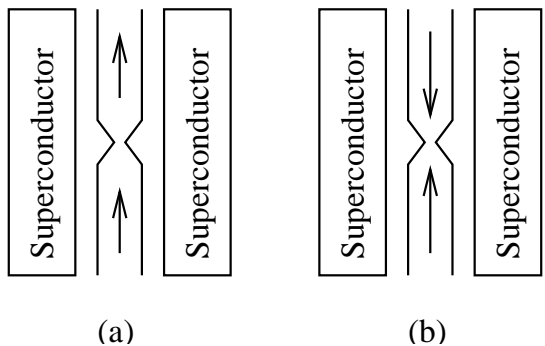}
\caption{The device considered in section~\ref{sec:out}.
In (a) there is no domain wall in the junction.
In (b) a domain wall is pinned in the junction.
}
\label{fig:schema4}
\end{figure}

\section{Sequential tunneling of Cooper pairs through a
magnetic domain wall}
\label{sec:out}

Now we consider the junction on
Fig.~\ref{fig:schema4} in which a ferromagnetic wire
is inserted in between two superconductors. In the absence
of a domain wall in the ferromagnetic wire (see
Fig.~\ref{fig:schema4}-(a)) the junction is just a 
S/F/S junction. In the presence of a domain wall
(see Fig.~\ref{fig:schema4}-(b)) Cooper pair-like states
arising from crossed Andreev reflection
can be transmitted through the junction.
As a consequence
the conductance is larger in
the presence of a magnetic domain wall. 
We consider two limiting cases:
\begin{itemize}
\item[(i)] Transport is dominated
by inelastic scattering in the ferromagnetic domains.
Because of inelastic scattering the distribution functions
in the ferromagnetic domains relax to the Fermi distribution.
This case is discussed in sections~\ref{sec:pert-tr-formu},
~\ref{sec:SDWN} and~\ref{sec:seq-SDWS}.
\item[(ii)] Transport through the domain wall is phase-coherent
and there is a Josephson current circulating between the two
ferromagnetic electrodes. This case is discussed in
section~\ref{sec:SDWS-Jo}.
\end{itemize}

\begin{figure}
\includegraphics [width=.4 \linewidth]{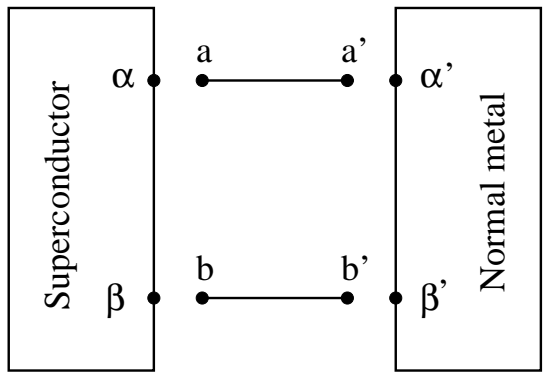}
\caption{The device considered in section~\ref{sec:SDWN}
in which two single channel electrodes representing two
magnetic domains are inserted in between a superconductor
and a normal metal. The two electrodes ending at sites
$a$ and $a'$ and sites $b$ and $b'$ are ferromagnetic.
In sections~\ref{sec:seq-SDWS} and~\ref{sec:SDWS-Jo}
we suppose that the electrode containing the sites
$\alpha'$ and $\beta'$ is superconducting.
}
\label{fig:schema5}
\end{figure}

\subsection{The different time scales}
\label{sec:dwell}
Similarly to Ref.~\cite{Brataas}
we notice that three time scales are involved
in out-of-equilibrium transport
through a ferromagnetic domain wall:
\begin{itemize}
\item[(i)] The 
transport dwell time $\tau_d$ being the time taken by
an electron to travel through one of the magnetic domains.
\item[(ii)] The energy relaxation time $\tau_E$.
Because of inelastic scattering the
distribution function in the out-of-equilibrium
conductor relaxes to the Fermi distribution. This
relaxation takes place on a time scale $\tau_E$.
\item[(iii)] The spin-flip time $\tau_{sf}$ being
the time above which spin-flip scattering
is relevant.
\end{itemize}
We suppose in this section that
$\tau_E \ll \tau_d \ll \tau_{sf}$. 
The distribution 
function in the intermediate magnetic
domains is thus well approximated by a
Fermi distribution. 
The chemical potential of
spin-up electrons is different from the
chemical potential of spin-down electrons.

\subsection{Perturbative transport formula}
\label{sec:pert-tr-formu}
In this section we discuss the perturbative transport
formula of the S/DW/N junction on Fig.~\ref{fig:schema5}.
The full transport formula to order $t^4$ is evaluated in
Appendix~\ref{app:tr-SDWN}.
The expression of $I_{a,\alpha}^{(\uparrow)}$
contains two kinds of terms: the
terms~(\ref{eq:I-a-alpha-1})~--~(\ref{eq:I-a-alpha-3})
describe processes taking place locally at the interfaces
between the superconductor and the ferromagnetic electrodes
without propagation in the ferromagnetic electrodes.
The terms~(\ref{eq:I-a-alpha-4})~--~(\ref{eq:I-a-alpha-6})
involve a propagation in the ferromagnetic electrodes.
The two kinds of terms would contribute if electrodes
$(a,a')$ and $(b,b')$ were single channel electrodes.
We consider here that electrodes $(a,a')$ 
and $(b,b')$ are multichannel electrodes and we
average the current
over the microscopic phases. Once this
averaging is done only
the ``local'' terms survive in the transport formula
given by 
\begin{eqnarray}
\label{eq:f-loc}
I_{a,\alpha}^{(\uparrow)} &=&
-4 \pi^2 t_\alpha^4 \tilde{\rho}_a^2
f_{\rm loc}^2 \left[ 1 - P_a^2 \right]
\left[ n_F(\omega-\mu_{a,\uparrow})
-n_F(\omega+\mu_{a,\downarrow}) \right]\\
\label{eq:g-alpha-beta}
&-&4 \pi^2 t_\alpha^2 t_\beta^2
\tilde{\rho}_a \tilde{\rho}_b \LL g_{\alpha,\beta}^2\RR
\left[ 1 + P_a \right] \left[ 1 + P_b \right]
\left[ n_F(\omega-\mu_{a,\uparrow})
-n_F(\omega-\mu_{b,\uparrow}) \right]\\
\label{eq:f-alpha-beta}
&-&4 \pi^2 t_\alpha^2 t_\beta^2 \tilde{\rho}_a
\tilde{\rho}_b \LL f_{\alpha,\beta}^2 \RR
\left[ 1 + P_a \right] \left[ 1 - P_b \right]
\left[ n_F(\omega-\mu_{a,\uparrow})
-n_F(\omega + \mu_{b,\downarrow}) \right]
.
\end{eqnarray}
The term~(\ref{eq:f-loc}) corresponds to
local Andreev reflection at the 
interface $a$~--~$\alpha$. The term~(\ref{eq:g-alpha-beta})
corresponds to elastic cotunneling through the superconductor
and the term~(\ref{eq:f-alpha-beta}) corresponds
to crossed Andreev reflection.

A similar calculation can be carried out at interface
$(a',\alpha')$. Once the average over the
microscopic phase variables is carried out we find
\begin{eqnarray}
\label{eq:terme1}
I_{a',\alpha'}^{(\uparrow)} &=&
-4 \pi^2 t_{\alpha'}^2 \tilde{\rho}_a 
\rho' \left[1+P_a\right]
\left[n_F(\omega-\mu_{a,\uparrow})
-n_F(\omega-\mu') \right]\\
\label{eq:terme2}
&+&8 \pi^4 t_{\alpha'}^4 \left(\tilde{\rho}_a\right)^2
\left(\rho'\right)^2 \left[1+P_a\right]^2
\left[ n_F(\omega-\mu_{a,\uparrow})
-n_F(\omega-\mu') \right]\\
\label{eq:terme3}
&+& 8 \pi^4 t_{\alpha'}^2 t_{\beta'}^2
\tilde{\rho}_a \tilde{\rho}_b
\LL \rho_{\alpha',\beta'}^2 \RR
\left[1+P_a\right] \left[1+P_b\right]
\left[ n_F(\omega-\mu_{b,\uparrow})
-n_F(\omega-\mu') \right]
,
\end{eqnarray}
where~(\ref{eq:terme1})
and~(\ref{eq:terme2}) describe electron tunneling
from the electrode $(a,a')$ into the normal metal
and~(\ref{eq:terme3}) describes elastic
cotunneling from electrode $(b,b')$ to electrode $(a,a')$.

\subsection{Sequential tunneling through the S/DW/N junction}
\label{sec:SDWN}
\label{sec:accu-SDWN}
In this section we discuss out-of-equilibrium transport
in a S/DW/N junction on the basis of the
two-channel model shown on
Fig.~\ref{fig:schema5}.  
We suppose that a voltage $V=0$ is applied on the superconductor
and a voltage $V'$ is applied on the normal metal.
The spin-up and spin-down
chemical potentials in the
two magnetic domains $(a,a')$ and $(b,b')$ are
determined in such a way that current is conserved.
In general there are four unknown chemical potentials
($\mu_{a,\uparrow}$, $\mu_{a,\downarrow}$,
$\mu_{b,\uparrow}$ and $\mu_{b,\downarrow}$) that
can be determined from four equations for current
conservation. There exist two cases in which the
$4\times 4$ system of equations can be reduced
to a $2 \times 2$ system of equations:
\begin{itemize}
\item[(i)] Half-metal ferromagnets where there
is only one spin population in each of the
ferromagnetic electrodes $(a,a')$ and $(b,b')$.
This case is treated in the main body of the
article.
\item[(ii)] The symmetric case where the two electrodes
$(a,a')$ and $(b,b')$ have identical density of states
and where $t_\alpha=t_\beta$ and $t_{\alpha'}
=t_{\beta'}$. This case is treated in
Appendix~\ref{app:sym}.
\end{itemize}

Let us consider half-metal ferromagnets:
$P_a=1$, $P_b=-1$. 
The transport formula is found to be
\begin{equation}
\label{eq:tr1}
\frac{I_{\rm tot}}{V'} = \frac{16 \pi^2}{\cal D}
t_\alpha^2 t_\beta^2 t_{\alpha'}^2 t_{\beta'}^2
\rho_{a,\uparrow} \rho_{b,\downarrow}
(\rho')^2 \LL f_{\alpha,\beta}^2 \RR
\left[1-2\pi^2 t_{\alpha'}^2 \rho_{a,\uparrow} \rho' \right]
\left[1-2\pi^2 t_{\beta'}^2 \rho_{b,\downarrow} \rho' \right]
,
\end{equation}
with
\begin{eqnarray}
{\cal D} &=& t_\alpha^2 t_\beta^2 t_{\alpha'}^2
\rho_{a,\uparrow} \rho' \LL f_{\alpha,\beta}^2 \RR
\left[1 - 2 \pi^2 t_{\alpha'}^2 \rho_{a,\uparrow} \rho' \right]
+
t_\alpha^2 t_\beta^2 t_{\beta'}^2 \rho_{b,\downarrow} \rho'
\LL f_{\alpha,\beta}^2\RR
\left[1 - 2\pi^2 t_{\beta'}^2 \rho_{b,\downarrow} \rho' \right]\\
\nonb
&+&t_{\alpha'}^2 t_{\beta'}^2 (\rho')^2
\left[1-2\pi^2 t_{\alpha'}^2 \rho_{a,\uparrow} \rho' \right]
\left[ 1 - 2 \pi^2 t_{\beta'}^2 \rho_{b,\downarrow} \rho' \right]
.
\end{eqnarray}

We note $\rho_N$ a typical value of the density of states
either in the superconductor or in the ferromagnetic
and normal metal electrodes. 
We first suppose
that $t_\alpha^2 \rho_N$ and  $t_\beta^2 \rho_N$
are small compared to $t_{\alpha'}$ and $t_{\beta'}$.
The transport formula takes the same form as in the
case where the ferromagnetic electrodes $(a,a')$
and $(b,b')$ are in equilibrium:
\begin{equation}
\label{eq:I-tot-P=1}
\frac{ I_{\rm tot}}{V'} = 16\pi^2 t_\alpha^2
t_\beta^2 \rho_{a,\uparrow} \rho_{b,\downarrow}
\LL f_{\alpha,\beta}^2 \RR
.
\end{equation}
In the other limiting case where $t_\alpha^2 \rho_N$ and 
$t_\beta^2 \rho_N$ are large compared to $t_{\alpha'}$ and
$t_{\beta'}$ we find
\begin{equation}
\label{eq:tggtp}
\frac{I_{\rm tot}}{V'} = 16 \pi^2 \rho'
\frac{ t_{\alpha'}^2 t_{\beta'}^2 \rho_{a,\uparrow}
\rho_{b,\downarrow}}{ t_{\alpha'}^2 \rho_{a,\uparrow}
+t_{\beta'}^2 \rho_{b,\downarrow} }
.
\end{equation}
We note $g_a=16 \pi^2 t_{\alpha'}^2 \rho_{a,\uparrow} \rho'$
and 
$g_b=16 \pi^2 t_{\alpha'}^2 \rho_{b,\downarrow} \rho'$ the
conductances associated to the interfaces $(a',\alpha')$ and
$(b',\beta')$. The total conductance is given
by $1/G_{\rm tot}=1/g_a+1/g_b$. The two
interfaces are thus in series which is
because transport is mediated by
crossed Andreev reflection: a spin-up electron from 
the normal metal is transfered at site $a'$, travels
to site $a$, is reflected as a spin-down hole at site $b$.
The spin-down hole travels to site $b'$ and is transfered
in the normal metal at site $\beta'$. As a consequence
of this transport process
the two interfaces $(a',\alpha')$ and $(b',\beta')$ are coupled
in series.

\subsection{Sequential tunneling through the S/DW/S junction}
\label{sec:seq-SDWS}
We consider 
the same model as in the preceding section but now
the electrode on the right is superconducting
(see Fig.~\ref{fig:schema5}).
We show that the properties of the S/DW/S junction are
similar to the properties of the S/DW/N junction.
We suppose that a voltage $V$ is applied on the left electrode
and a voltage $V'$ is applied on the right electrode.
We consider a regime in which
inelastic scattering in the ferromagnetic electrodes
is strong enough so that
the transport dwell time is much larger than
the energy relaxation time (see section~\ref{sec:dwell}).
Moreover we suppose that inelastic scattering is strong
enough so that there is no Josephson effect.

We consider that the ferromagnetic electrodes $(a,a')$
and $(b,b')$ are half-metal ferromagnets: $P_a=1$, $P_b=-1$.
The total current is given by
\begin{equation}
\label{eq:Itot-half-SS}
\frac{I_{\rm tot}}{V'-V} =
16 \pi^2 \rho_{a,\uparrow} \rho_{b,\downarrow}
\frac{ t_\alpha^2 t_\beta^2 t_{\alpha'}^2 t_{\beta'}^2
\LL f_{\alpha,\beta}^2 \RR 
\LL f_{\alpha',\beta'}^2 \RR }
{t_\alpha^2 t_\beta^2 \LL f_{\alpha,\beta}^2 \RR
+t_{\alpha'}^2 t_{\beta'}^2 
\LL f_{\alpha',\beta'}^2 \RR }
.
\end{equation}
If we note $g=16\pi^2 t_\alpha^2 t_\beta^2
\LL f_{\alpha,\beta}^2 \RR$ and
$g'=16\pi^2 t_{\alpha'}^2 t_{\beta'}^2
\LL f_{\alpha',\beta'}^2 \RR$ the conductances associated
to crossed Andreev reflection at the contacts
with each of the two superconductors we see that the total
conductance is such that
\begin{equation}
\label{eq:series-SS}
\frac{1}{G_{\rm tot}} = \frac{1}{g} + \frac{1}{g'}
.
\end{equation}
which is the expected result since electron pairs travel
in series through the two superconductors.

\section{Josephson effect in the S/DW/S Josephson junction}
\label{sec:SDWS-Jo}

\subsection{A two-channel model}

Now we consider the S/DW/S junction on Fig.~\ref{fig:schema5}
in which the right electrode is superconducting. 
We suppose that $\tau_E \sim \tau_d \ll \tau_{sf}$ or
$\tau_E < \tau_d \ll \tau_{sf}$ so that the two
superconductors can be coupled coherently through the
two ferromagnetic channels. We look for the condition under
which a Josephson current can circulate across the junction.
We suppose in this section
that the two electrodes $(a,a')$ and $(b,b')$
are half-metal ferromagnets with antiparallel spin
orientations. The case of a partial spin
polarization and non collinear spin orientations will
be discussed in section~\ref{sec:multichannel}.

The Nambu representation of the hopping matrix elements is given by
\begin{eqnarray}
\Ht_{\alpha,a} &=& \left[ \begin{array}{cc}
t_\alpha e^{-i (\varphi-\chi)/4} & 0 \\
0 & -t_\alpha e^{i(\varphi-\chi)/4} \end{array} \right]\\
\Ht_{\beta,b} &=& \left[ \begin{array}{cc}
t_\beta e^{-i(\varphi+\chi)/4} & 0 \\
0 & -t_\beta e^{i(\varphi+\chi)/4} \end{array} \right]\\
\Ht_{a',\alpha'} &=& \left[ \begin{array}{cc}
t_{\alpha'} e^{-i (\varphi-\chi)/4} & 0 \\
0 & -t_{\alpha'} e^{i(\varphi-\chi)/4} \end{array} \right]\\
\Ht_{b',\beta'} &=& \left[ \begin{array}{cc}
t_{\beta'} e^{-i(\varphi+\chi)/4} & 0 \\
0 & -t_{\beta'} e^{i(\varphi+\chi)/4} \end{array} \right]
,
\end{eqnarray}
where $\varphi$ is the difference between the superconducting
phases in the right and left electrode and $\chi$ is the
magnetic flux through the loop.
We have the relations
$\Ht_{a,\alpha}=\left(\Ht_{\alpha,a}\right)^*$,
$\Ht_{b,\beta}=\left(\Ht_{\beta,b}\right)^*$,
$\Ht_{\alpha',a'}=\left(\Ht_{a',\alpha'}\right)^*$
and
$\Ht_{\beta',b'}=\left(\Ht_{b',\beta'}\right)^*$.
The equilibrium
current flowing from site~$\alpha$ to site~$a$ is given by
\begin{equation}
\label{eq:super-eq}
I_{\alpha,a} = \frac{e}{h} \int d\omega
n_F(\omega) \mbox{Tr} \left\{ \hat{\sigma}^z \left[
\Ht_{\alpha,a} \left( \HG^A_{a,\alpha}-\HG^R_{a,\alpha} \right)
-\Ht_{a,\alpha} \left( \HG^A_{\alpha,a} - \HG^R_{\alpha,a} \right)
\right] \right\}
.
\end{equation}
The Green's functions are $2 \times 2$ matrices
since we do not discuss non collinear magnetizations for the moment.

We deduce from (\ref{eq:super-eq}) that to order $t^4$
the spin-up current
through electrode $(a,a')$ is given by
\be
\label{eq:super-SDWS}
I_{\alpha,a}^{(\uparrow)} =
-2 i \frac{e}{h} t_\alpha t_\beta t_{\alpha'} t_{\beta'}
\sin{\varphi} \int_0^{+\infty} d\omega
f_{\alpha,\beta}(\omega) f_{\alpha',\beta'}(\omega)
\left\{ \frac{ g_{a,a'}^{A,11} g_{b,b'}^{A,22} }
{ \mbox{Det}[\CHI-\CHK^A] }
-\frac{g_{a,a'}^{R,11} g_{b,b'}^{R,22} }
{\mbox{Det} [\CHI-\CHK^R] } \right\}
,
\ee
where $\CHK^{A,R}$ is the $4 \times 4$
matrix involved in the Dyson
equation $[\CHI-\CHK^{A,R} ] \CHG^{A,R}=\CHg^{A,R}$:
\be
\label{eq:K4x4}
\left[ \begin{array}{cccc}
1-K_{a,a}^{1,1} & -K_{b,a}^{2,1} & -K_{a',a}^{1,1} & -K_{b',a}^{2,1} \\
-K_{a,b}^{1,2} & 1-K_{b,b}^{2,2} & -K_{a',b}^{1,2} & -K_{b',b}^{2,2} \\
-K_{a,a'}^{1,1} & -K_{b,a'}^{2,1} & 1-K_{a',a'}^{1,1} & -K_{b',a'}^{2,1} \\
-K_{a,b'}^{1,2} & -K_{b,b'}^{2,2} & -K_{a',b'}^{1,2} & 1-K_{b',b'}^{2,2} 
\end{array} \right]
\left[ \begin{array}{c}
G_{a,a}^{1,1} \\ G_{a,b}^{1,2} \\ G_{a,a'}^{1,1} \\ G_{a,b'}^{1,2}
\end{array} \right]
= \left[ \begin{array}{c}
g_{a,a}^{1,1} \\ 0 \\ g_{a,a'}^{1,1} \\ 0
\end{array} \right]
,
\ee
where we used the notation $K_{a,a}^{1,1}
=t_{a,\alpha}^{1,1} g_{\alpha,\alpha}
t_{\alpha,a}^{1,1} g_{a,a}^{1,1}$,
$K_{b,a}^{2,1} = t_{b,\beta}^{2,2}
f_{\beta,\alpha} t_{\alpha,a}^{1,1} g_{a,a}^{1,1}$, etc.
The role of disorder can be included in a straightforward fashion.
Since the spin-up and spin-down electrons of the Cooper pair
propagate in different electrodes we should replace
$g_{a,a'}^{A,1,1}$ and $g_{b,b'}^{A,2,2}$ by their average
over disorder which decay exponentially with distance over a length
scale equal to the elastic mean free path~\cite{Abrikosov}. We conclude
that a Josephson current cannot be observed under usual experimental
conditions since the size of the ferromagnetic region is usually
much larger than the elastic mean free path. The opposite limit of
small disorder is considered in Appendix~\ref{app:small-dis}.

\subsection{Multichannel effects}
\label{sec:multichannel}
\label{sec:multi}
\begin{figure}
\includegraphics [width=.3 \linewidth]{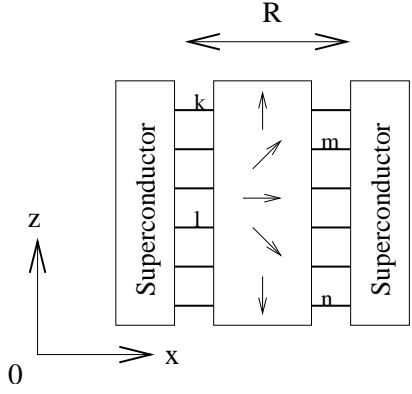}
\caption{Schematic representation of the Josephson junction
containing a domain wall with a rotating magnetization.
The $x$- and $z$-axis are shown on the figure. The $y$-axis
is perpendicular to the figure.
}
\label{fig:Jo-DW}
\end{figure}

\subsubsection{Transport formula}

We consider 
in this section the multichannel S/DW/S junction on
Fig.~\ref{fig:Jo-DW} in which the ferromagnetic metal
is multi-connected to the superconductors.
The local magnetization makes an angle $\theta(z)$ with the $z$-axis. The 
situation with a uniform $\theta$ corresponds to the multichannel 
$\pi$-junction.

The supercurrent
is given by
\be
I_S = \frac{1}{2} \frac{e}{h} \int_0^{+\infty}
d \omega \sum_n \mbox{Tr} \left\{
\hat{\sigma}^z \left[
\Ht_{\alpha_n,a_n} \left( \HG^A_{a_n,\alpha_n}
-\HG^R_{a_n,\alpha_n} \right)
-\Ht_{a_n,\alpha_n} \left( \HG^A_{\alpha_n,a_n}
-\HG^R_{\alpha_n,a_n} \right) \right] \right\}
,
\ee
where the Green's functions are $4 \times 4$ matrices.
The supercurrent to order $t_\alpha^2 t_\beta^2$
can be written as
\begin{eqnarray}
\label{eq:66}
I_S &=& -2i \frac{e}{h}
t_\alpha^2 t_\beta^2 \sin{\varphi}
\int_0^{+\infty}
d\omega \sum_{k,l,m,n}
f_{\alpha_m,\alpha_n} f_{\beta_k,\beta_l}\\\nonb
&\times&
\left\{ 
\left[
\frac{g_{a_m,b_k}^{2,2,A} g_{a_n,b_l}^{1,1,A}}
{\mbox{Det}[\CHI-\CHK^A]}
-
\frac{g_{a_m,b_k}^{2,2,R} g_{a_n,b_l}^{1,1,R}}
{\mbox{Det}[\CHI-\CHK^R]} \right]
+\left[ \frac{g_{a_m,b_k}^{4,4,A} g_{a_n,b_l}^{3,3,A}}
{\mbox{Det}[\CHI-\CHK^A]}
- \frac{g_{a_m,b_k}^{4,4,R} g_{a_n,b_l}^{3,3,R}}
{\mbox{Det}[\CHI-\CHK^R]} \right]
+2 \left[
\frac{ g_{a_m,b_k}^{2,4,A} g_{a_n,b_l}^{1,3,A} }
{\mbox{Det}[\CHI-\CHK^A]}
-\frac{ g^{2,4,R}_{a_m,b_k} g^{1,3,R}_{a_n,b_l} }
{\mbox{Det}[\CHI-\CHK^R]}\right]
\right\}
,
\end{eqnarray}
where $g_{a_k,b_m}^{i,i}$ is
$i$-th Nambu component of the
propagator connecting the two ends of the
ferromagnetic metal at sites $a_k$ and $b_m$.
The first two terms in the r.h.s. of Eq.~(\ref{eq:66})
correspond to a propagation without spin-flip in the 
ferromagnetic region whereas the last term correspond to
a propagation with spin flip.
We deduce from Eq.~(\ref{eq:66}) the same conclusions
as in the two-channel model. Namely an average Josephson
current can circulate only if the size of the ferromagnetic
region is smaller than the elastic mean free path in
the ferromagnetic metal, a condition that is not
usually realized in experiments. 

\subsubsection{Limit of small disorder}

\begin{figure}
\includegraphics [width=.4 \linewidth]{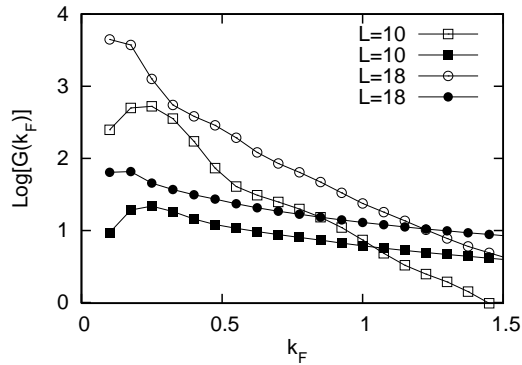}
\caption{Variation of $\log_{10}{[G(k_F)]}$ as a function of
$k_F$. $k_F$ is measured in units of $1/a_0$.
We restrict the summation in (\ref{eq:GkF})
to $-L_z/2 \le z_n, z_l <0$, $0<z_m,z_k \le L_z/2$,
$-L_y/2 \le y_n,y_l,y_m,y_k \le L_y/2$ and we use
$L=L_y=L_z$. The open symbols correspond to all
trajectories in (\ref{eq:GkF}). The filled
symbols correspond only to the trajectories 
perpendicular to the interface.
}
\label{fig:G2}
\end{figure}

To obtain the supercurrent in the ballistic limit
we replace the propagators by their expressions
and sum over all channels. 
The propagators
of a ferromagnetic metal with a rotating magnetization are not
known in general. This is why we discuss here only the situation
where the width of the domain wall is vanishingly small
and the ferromagnets are half-metal ferromagnets. In
this case there is no spin precession
in the ferromagnetic region but there exist
trajectories parallel to the interface that we can take into
account in our calculation.
The supercurrent is given by
\be
I_S = 8 \pi \frac{e}{h} L_y t_\alpha^2 t_\beta^2 \sin{\varphi}
\left(\frac{m a_0^2}{\hbar^2}\right)^4 
G(k_F)
,
\ee
with
\begin{eqnarray}
\label{eq:GkF}
G(k_F) &=& \frac{1}{L_y}
\int_{-\infty}^0 \frac{d z_n}{a_0}
\int_{-\infty}^{+\infty} \frac{d y_n}{a_0}
\int_{-\infty}^0 \frac{d z_l}{a_0}
\int_{-\infty}^{+\infty} \frac{d y_l}{a_0}
\int_0^{+\infty} \frac{d z_m}{a_0}
\int_{-\infty}^{+\infty} \frac{d y_m}{a_0}
\int_0^{+\infty} \frac{d z_k}{a_0}
\int_{-\infty}^{+\infty} \frac{d y_k}{a_0}\\
&& \frac{a_0}{2\pi R_{\alpha_m,\alpha_n}}
\frac{a_0}{2\pi R_{\beta_k,\beta_l}}
\frac{a_0}{2\pi R_{a_m,b_k}}
\frac{a_0}{2\pi R_{a_n,b_l}}
\sin{\left[k_F R_{\alpha_m,\alpha_n}\right]}
\sin{\left[k_F R_{\beta_k,\beta_l}\right]}
\cos{\left[k_F \left(R_{a_m,b_k}-R_{a_n,b_l}\right)\right]}
,
\end{eqnarray}
and with
\begin{eqnarray}
R_{\alpha_m,\alpha_n} &=& \sqrt{ (z_n-z_m)^2+(y_m-y_n)^2} \\
R_{\beta_k,\beta_l} &=& \sqrt{ (z_k-z_l)^2+(y_k-y_l)^2} \\
R_{a_m,b_k} &=& \sqrt{ R^2+(z_m-z_k)^2+(y_m-y_k)^2} \\
R_{a_n,b_l} &=& \sqrt{ R^2+(z_n-z_l)^2+(y_n-y_l)^2} 
,
\end{eqnarray}
where $R$ is the longitudinal dimension of the junction
(see Fig.~\ref{fig:Jo-DW}).
We have shown on Fig.~\ref{fig:G2} the variation of 
$G(k_F)$ as a function of $k_F$. We see that strong finite size
effects are present but still we can make a comparison
between (i) a calculation in which all trajectories are taken
into account and (ii) a calculation in which only the
trajectories perpendicular to the interface are taken into
account.
We see that for small values of $k_F$ (typically 
$k_F$ smaller than $1/a_0$, where $a_0$ is the lattice
parameter) the summation (i)
is larger than (ii) whereas the opposite is true for
larger values of $k_F$. This shows that trajectories
parallel to the interface play a relevant role in
the determination of the supercurrent.

\section{Conclusions}
\label{sec:conclu}

To conclude we have presented a detailed investigation of
several mechanisms involved in transport across several
junctions involving ferromagnetic domain walls
(S/DW, S/DW/N and S/DW/S junctions)
The role of non collinear magnetization was
studied for the S/DW junctions. 
Using a spin $\otimes$ Nambu $\otimes$ Keldysh formalism
we have derived the form of lowest order transport formula
valid for an arbitrary profile of magnetization. We find
that the conductance is a scaling function of
$\xi_0/D$, where $\xi_0$ is the zero-energy BCS correlation
length and $D$ is the width of the domain wall.
Because of the proximity effect an exchange field
can be induced in the superconductor. Neglecting the spatial
variation of the exchange field, we have derived
the transport formula and shown that there was no spin precession
around the axis of the exchange field. 
We discussed the transport formula of the
S/DW/N junction. We have shown that to lowest order only the
processes taking place locally at each interface played a role.
These processes are: elastic cotunneling
through the superconductor, crossed Andreev reflection,
electron tunneling from the ferromagnet to the normal metal and
elastic cotunneling through the normal metal.
We described the transport of Cooper pairs across
the S/DW/N and
S/DW/S junctions in a regime where transport is dominated
by inelastic scattering but spin is conserved. With these
assumptions the local distribution function within the domain
wall is a Fermi distribution with a different spin-up and
spin-down chemical potential. This
model provides a detailed description of the sequential
tunneling of Cooper pairs across the S/DW/N and S/DW/S
junctions.
We described the Josephson effect in a
S/DW/S junction. 
Diffusion is usually
strong in a ferromagnet and disorder is thus expected to play
a relevant role. In particular the Josephson current decays
exponentially with the longitudinal dimension of the junction.
The characteristic length is equal to
the mean free path in the ferromagnetic metal.
This means that a Josephson current cannot be observed in usual
conditions. Nevertheless there can exist a finite current due
to crossed Andreev reflection associated to elastic cotunneling
in the ferromagnetic region.

\section*{Acknowledgements}
The authors acknowledge fruitful discussions with
H. Courtois, D. Feinberg M. Giroud and B. Pannetier.


\appendix
\section{Spin precession in the metallic case}
\label{app:precession}

\begin{figure}
\includegraphics [width=.4 \linewidth]{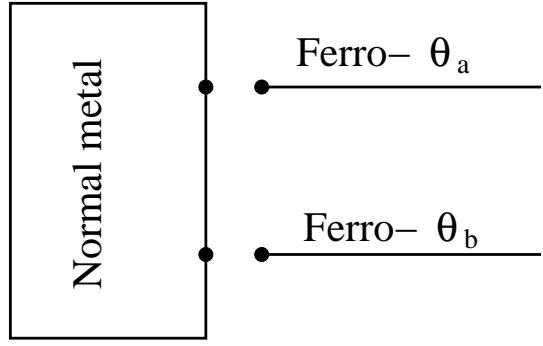}
\caption{The device considered in appendix~\ref{app:precession}.
Electrode ending at site ``a''
is a ferromagnet with a magnetization pointing in
the direction $\theta_a$.
Electrode ending at site ``b'' is a 
ferromagnet with a magnetization pointing
in the direction $\theta_b$.
}
\label{fig:schema3}
\end{figure}

In this appendix we consider the junction on
Fig.~\ref{fig:schema3} in which two ferromagnetic electrodes
with non collinear magnetizations
are connected to a normal metal~\cite{Hernando,Jedema}.
Our goal is to provide
a comparison with the superconducting case
presented in section~\ref{sec:precess}. 
We suppose that a magnetic
field $h$ is applied on the normal metal and that
the only effect of the magnetic field is
to generate Zeeman splitting. 
The crossed conductance $G_{a,b}=\partial
I_a / \partial V_b$ associated to elastic cotunneling
takes the form
\begin{eqnarray}
G_{a,b} &=& 8 \pi^2 t_\alpha^2 t_\beta^2
\tilde{\rho}_a \tilde{\rho}_b
\left( \frac{m a_0^2}{\hbar^2} \right)^2
\left( \frac{a_0}{2\pi R_{\alpha,\beta}} \right)^2
\left\{ 1 + P_a P_b \cos{\theta_a} \cos{\theta_b} \right.\\\nonb
&+& \left. P_a P_b \sin{\theta_a} \sin{\theta_b}
\cos{ \left\{ \left[ k_{F,\uparrow}
-k_{F,\downarrow} \right] R_{\alpha,\beta}
+e V_b \left[ \frac{1}{v_{F,\uparrow}}
-\frac{1}{v_{F,\downarrow}} \right]
R_{\alpha,\beta} \right\}} \right\}
.
\end{eqnarray}
Spin precession can have two origins:
(i) the term $\left[k_{F,\uparrow}-k_{F,\downarrow}\right]
R_{\alpha,\beta}$ describes oscillations of the conductance
due to a mismatch in the Fermi wave-vectors; (ii) the
term $\left[ \frac{1}{v_{F,\uparrow}}
-\frac{1}{v_{F,\downarrow}} \right]
R_{\alpha,\beta}$ describes oscillations in the conductance
due to a mismatch in the Fermi velocities. 

\section{Perturbative transport formula of the two channel
S/DW/N junction}
\label{app:tr-SDWN}
In this appendix we provide a derivation of the transport
formula of the S/DW/N model represented
on Fig.~\ref{fig:schema5}. 

\subsection{Transport at interface $(a,\alpha)$}

The current through each link of the network
on Fig.~\ref{fig:schema5} is
given by the transport formula~(\ref{eq:transport-formula}).
The spin-up current through the link $\alpha$~--~$a$
is found to be
\begin{eqnarray}
\label{eq:I-a-alpha-1}
I_{a,\alpha}^{(\uparrow)} &=&
-4 \pi^2 t_\alpha^4 \tilde{\rho}_a^2
f_{\rm loc}^2 \left[ 1 - P_a^2 \right]
\left[ n_F(\omega-\mu_{a,\uparrow})
-n_F(\omega+\mu_{a,\downarrow}) \right]\\
\label{eq:I-a-alpha-2}
&-&4 \pi^2 t_\alpha^2 t_\beta^2
\tilde{\rho}_a \tilde{\rho}_b g_{\alpha,\beta}^2
\left[ 1 + P_a \right] \left[ 1 + P_b \right]
\left[ n_F(\omega-\mu_{a,\uparrow})
-n_F(\omega-\mu_{b,\uparrow}) \right]\\
\label{eq:I-a-alpha-3}
&-&4 \pi^2 t_\alpha^2 t_\beta^2 \tilde{\rho}_a
\tilde{\rho}_b f_{\alpha,\beta}^2
\left[ 1 + P_a \right] \left[ 1 - P_b \right]
\left[ n_F(\omega-\mu_{a,\uparrow})
-n_F(\omega + \mu_{b,\downarrow}) \right]\\
\label{eq:I-a-alpha-4}
&-&4 \pi^2 t_\alpha^2 t_{\alpha'}^2
\rho_{a,a'}^\uparrow \rho_{a,a'}^\downarrow
f_{\rm loc}^2 \left[ n_F(\omega-\mu_{a,\uparrow})
-n_F(\omega+\mu_{a,\downarrow}) \right]\\
\label{eq:I-a-alpha-4bis}
&-&4 \pi^2 t_\alpha t_\beta t_{\alpha'} t_{\beta'}
\rho_{a,a'}^\uparrow \rho_{b,b'}^\uparrow
g_{\alpha,\beta} g_{\alpha',\beta'}
\left[ n_F(\omega-\mu_{a,\uparrow})
-n_F(\omega-\mu_{b,\uparrow}) \right]\\
\label{eq:I-a-alpha-5}
&-&4\pi^2 t_\alpha t_\beta t_{\alpha'}
t_{\beta'} \rho_{a,a'}^\uparrow
\rho_{b,b'}^\downarrow 
f_{\alpha,\beta} f_{\beta,\alpha}
\left[ n_F(\omega-\mu_{a,\uparrow})
-n_F(\omega+\mu_{b,\downarrow}) \right]\\
\label{eq:I-a-alpha-6}
&-&4\pi t_\alpha t_\beta t_{\alpha'} t_{\beta'}
\mbox{Im} \left[ g_{a,a'}^{\uparrow R}
g_{b,b'}^{\uparrow A} \right]
g_{\alpha,\beta} g_{\alpha',\beta'}
n_F(\omega-\mu')
,
\end{eqnarray}
where $\mu_{a,\uparrow}$ and $\mu_{a,\downarrow}$
are the spin-up and spin-down chemical potentials
in electrode $(a,a')$,
$\mu_{b,\uparrow}$ and $\mu_{b,\downarrow}$
are the spin-up and spin-down chemical potentials
in electrode (b,b'), and $\mu'$ is the
chemical potential in the normal metal.
After phase averaging we obtain three 
contributions to the transport formula:
local Andreev reflection given by (\ref{eq:f-loc}),
elastic cotunneling through the superconductor
given by (\ref{eq:g-alpha-beta}) and
crossed Andreev reflection given by (\ref{eq:f-alpha-beta}).

\subsection{Transport at interface $(a',\alpha')$}
\label{app:aprime}
The same calculation can be carried out at interface
$(a',\alpha')$.
The transport formula is found to be
\begin{eqnarray}
I_{a',\alpha'}^{(\uparrow)} &=&
4 \pi^2 t_{\alpha'}^2 t_{\beta'}^2
\tilde{\rho}_a \tilde{\rho}_b
g^A_{\alpha',\beta'} g^R_{\alpha',\beta'}
\left[ 1 + P_a \right] \left[ 1 + P_b \right]
n_F(\omega-\mu_{b,\uparrow})\\\nonb
&-& 8 \pi^4 t_{\alpha'}^2 t_{\beta'}^2
\tilde{\rho}_a \tilde{\rho}_b
\rho_{\alpha',\beta'}^2 \left[1+P_a\right]
\left[1+P_b\right] n_F(\omega-\mu')\\\nonb
&-& 4 \pi^2 t_{\alpha'}^2 \tilde{\rho}_a
\rho' \left[1+P_a\right]
\left[ n_F(\omega-\mu_{a,\uparrow})
-n_F(\omega-\mu') \right]\\\nonb
&+& 8 \pi^4 t_{\alpha'}^4
\left( \tilde{\rho}_a \right)^2
\left( \rho' \right)^2
\left[1+P_a\right]^2
\left[ n_F(\omega-\mu_{a,\uparrow})
-n_F(\omega-\mu') \right] \\\nonb
&-& 8 \pi^2 t_\alpha^2 t_{\alpha'}^2
\rho' g \mbox{Re}\left[ g_{a,a'}^\uparrow\right]
\rho_{a,a'}^\uparrow n_F(\omega-\mu_{a,\uparrow})\\\nonb
&+&2 \pi^2 t_\alpha^2 t_{\alpha'}^2
\left( \rho' \right)^2
\left( g_{a,a'}^{\uparrow R} \right)^2
n_F(\omega-\mu')\\\nonb
&-& 2 i \pi t_{\alpha}^2 t_{\alpha'}^2
\rho' g \left( g_{a,a'}^{\uparrow A} \right)^2
n_F(\omega-\mu')\\\nonb
&-&4 \pi^2 t_\alpha t_\beta t_{\alpha'} t_{\beta'}
g_{\alpha,\beta} \rho_{a,a'}^\uparrow
\mbox{Im} \left[ g_{\alpha',\beta'}^A g_{b,b'}^{\uparrow A}
\right] n_F(\omega-\mu_{a,\uparrow})\\\nonb
&+&2 i \pi t_\alpha t_\beta t_{\alpha'} t_{\beta'}
g_{\alpha,\beta} \rho_{b,b'}^\uparrow
\left[ g_{\alpha',\beta'}^A g_{a,a'}^{\uparrow R}
+i\pi \rho' g_{a,a'}^{\uparrow A} \right]
n_F(\omega-\mu_{b,\uparrow})\\\nonb
&+&2 i \pi t_\alpha t_\beta t_{\alpha'}
t_{\beta'} g_{\alpha,\beta} \rho_{\alpha',\beta'}
\left[ g_{a,a'}^{\uparrow R} g_{b,b'}^{\uparrow R}
-i\pi \tilde{\rho}_b (1+P_b) g_{a,a'}^{\uparrow A}
\right] n_F(\omega-\mu')\\\nonb
&-&4\pi^2 t_{\alpha'}^2 t_{\beta'}^2
\tilde{\rho}_a \tilde{\rho}_b
\mbox{Re} \left[ \left( g_{\alpha',\beta'}^A
\right)^2 \right] \left[1+P_a\right]
\left[1+P_b\right] n_F(\omega-\mu_{a,\uparrow})
.
\end{eqnarray}
After averaging over the phase variables
we obtain the transport
formula given by Eqs.~(\ref{eq:terme1})~-- (\ref{eq:terme3})
that contains only two processes: tunneling from
site $a'$ to site $\alpha'$ and elastic cotunneling
from site $b'$ to site $a'$.

\section{Transport formula of the symmetric two-channel
S/DW/N and S/DW/S junctions}
\label{app:sym}

In this appendix we consider S/DW/N and S/DW/S junctions
with two symmetric channels. With this model we confirm
the results obtained in the main body of the article
for the asymmetric junction with half-metal ferromagnets.
We suppose that the two channels have an identical
density of states: $\tilde{\rho}_a=\tilde{\rho}_b$,
that the tunnel matrix elements are identical in the
two channels: $t=t_\alpha=t_\beta$,
$t'=t_{\alpha'}=t_{\beta'}$, and that the two channels
have an opposite spin polarization: $P_a=P$ and $P_b=-P$.
Then there exists a simple symmetry relation between
the chemical potentials in the two ferromagnetic electrodes:
$\mu_{a,\uparrow}=\mu_{b,\downarrow}$ and
$\mu_{a,\downarrow}=\mu_{b,\uparrow}$. 

\subsection{The S/DW/N junction}

In the limiting case $t \ll t'$ we have
$\mu_{a,\uparrow} \simeq V'$.
The transport formula is identical to the case where
electrodes $(a,a')$ and $(b,b')$ are in equilibrium:
\begin{equation}
\label{eq:I-tot-P}
\frac{ I_{\rm tot} }{V'} =
32 \pi^2 t^4 \tilde{\rho}^2 (1-P^2)
f_{\rm loc}^2 
+32 \pi^2 t^4 \tilde{\rho}^2 (1+P^2)
\LL f_{\alpha,\beta}^2 \RR
.
\end{equation}
In the case of half-metal ferromagnets ($P=1$)
only the term corresponding to crossed Andreev reflection is
non-zero and Eq.~(\ref{eq:I-tot-P}) is equivalent to 
Eq.~(\ref{eq:I-tot-P=1}).

In the limiting case $\rho_N t' \ll \rho_N t \ll 1$ and
$t' \ll \rho_N t^2$ the current is the sum of a
contribution due to local
Andreev reflection and a contribution due to crossed
Andreev reflection: $I_{\rm tot}=I_{\rm AR}
+I_{\rm CAR}$, with
\begin{eqnarray}
\label{eq:IAR-P}
\frac{I_{\rm AR}}{V'} &=& 
\frac{32 \pi^2 (t')^2 \tilde{\rho} \rho' (1-P^2)
f_{\rm loc}^2 \left( \LL f_{\alpha,\beta}^2 \RR
+\LL g_{\alpha,\beta}^2\RR \right)}
{ (1-P^2) \left[ f_{\rm loc}^2 \LL g^2_{\alpha,\beta} \RR
+\LL f^2_{\alpha,\beta} \RR^2 \right]
+(1+P^2) \LL f^2_{\alpha,\beta} \RR \left[
f_{\rm loc}^2 + \LL g^2_{\alpha,\beta} \RR \right]}\\
\frac{I_{\rm CAR}}{V'} &=& 
\frac{16 \pi^2 (t')^2 \tilde{\rho} \rho' 
\LL f^2_{\alpha,\beta} \RR \left[
(1+P^2) \left( \LL f^2_{\alpha,\beta} \RR
+\LL g^2_{\alpha,\beta} \RR \right)
+2 P^2 \left( f_{\rm loc}^2 -
\LL f^2_{\alpha,\beta} \RR \right) \right]}
{(1-P^2) \left[ f_{\rm loc}^2 \LL g^2_{\alpha,\beta} \RR
+\LL f^2_{\alpha,\beta} \RR^2 \right]
+(1+P^2) \LL f^2_{\alpha,\beta} \RR \left[
f_{\rm loc}^2 + \LL g^2_{\alpha,\beta} \RR \right]}
\label{eq:ICAR-P}
.
\end{eqnarray}
In the case of half-metal ferromagnets~(\ref{eq:IAR-P})
and~(\ref{eq:ICAR-P}) are equivalent to~(\ref{eq:tggtp}).

\subsection{The S/DW/S junction}

In the case of the S/DW/S junction
the total current is the sum of the local Andreev reflection
and crossed Andreev reflection terms:
\begin{eqnarray}
\label{eq:IAR-SDWS}
\frac{I_{AR}}{V'-V} &=&
\frac{128 \pi^2}{\cal D'} t^4 (t')^4 \tilde{\rho}^4
f_{\rm loc}^2 (1-P^2) \left\{
\left[ t^4 \LL g^2_{\alpha,\beta} \RR +
(t')^4 \LL g^2_{\alpha',\beta'} \RR \right]
\left[ (1-P^2) f_{\rm loc}^2 +
(1+P^2) \LL f^2_{\alpha',\beta'} \RR \right]\right.\\\nonb
&+& \left.
\left[ t^4 \LL f^2_{\alpha,\beta} \RR +
(t')^4 \LL f^2_{\alpha',\beta'} \RR \right]
\left[ (1+P^2) f_{\rm loc}^2 +
(1-P^2) \LL f^2_{\alpha',\beta'} \RR \right]
\right\}\\
\label{eq:ICAR-SDWS}
\frac{I_{CAR}}{V'-V} &=&
\frac{128 \pi^2}{\cal D'} t^4 (t')^4 \tilde{\rho}^4
\LL f_{\alpha,\beta}^2 \RR (1+P^2) \left\{
\left[ t^4 \LL g^2_{\alpha,\beta} \RR +
(t')^4 \LL g^2_{\alpha',\beta'} \RR \right]
\left[ (1-P^2) f_{\rm loc}^2 +
(1+P^2) \LL f^2_{\alpha',\beta'} \RR \right]\right.\\\nonb
&+& \left.
\left[ t^4 \LL f^2_{\alpha,\beta} \RR +
(t')^4 \LL f^2_{\alpha',\beta'} \RR \right]
\left[ (1+P^2) f_{\rm loc}^2 +
(1-P^2) \LL f^2_{\alpha',\beta'} \RR \right]
\right\}
,
\end{eqnarray}
with
\begin{eqnarray}
{\cal D}' &=& 4 t^8 \tilde{\rho}^2 \left\{
(1-P^2) \left[ f_{\rm loc}^2 \LL g^2_{\alpha,\beta} \RR
+\LL f^2_{\alpha,\beta}\RR^2 \right]
+(1+P^2) \left[ f_{\rm loc}^2 \LL f^2_{\alpha,\beta} \RR
+\LL f^2_{\alpha,\beta} \RR \LL g^2_{\alpha,\beta} \RR
\right] \right\}\\\nonb
&+& 4 (t')^8 \tilde{\rho}^2 \left\{
(1-P^2) \left[ f_{\rm loc}^2 \LL g^2_{\alpha',\beta'} \RR
+\LL f^2_{\alpha',\beta'} \RR^2 \right]
+(1+P^2) \left[ f_{\rm loc}^2 \LL f^2_{\alpha',\beta'} \RR^2
+\LL f^2_{\alpha',\beta'} \RR \LL g^2_{\alpha',\beta'} \RR
\right] \right\}\\\nonb
&+& 4 t^4 (t')^4 \tilde{\rho}^2 \left\{
(1-P^2) \left[ f_{\rm loc}^2 \LL g^2_{\alpha,\beta} \RR
+f_{\rm loc}^2 \LL g^2_{\alpha',\beta'} \RR
+2 \LL f^2_{\alpha,\beta} \RR \LL f^2_{\alpha',\beta'} \RR
\right] \right.\\\nonb
&+&\left.(1+P^2) \left[ f_{\rm loc}^2 \LL f^2_{\alpha,\beta} \RR
+f_{\rm loc}^2 \LL f^2_{\alpha',\beta'} \RR
+\LL f^2_{\alpha,\beta} \RR \LL g^2_{\alpha',\beta'} \RR
+\LL f^2_{\alpha',\beta'} \RR \LL g^2_{\alpha,\beta} \RR
\right] \right\}
.
\end{eqnarray}
If the contacts with the two superconductors are identical
we have $t=t'$, $f_{\alpha,\beta}=f_{\alpha',\beta'}$
and $g_{\alpha,\beta}=g_{\alpha',\beta'}$ from what we
deduce
\begin{eqnarray}
\label{eq:IAR-sym-SS}
\frac{I_{AR}}{V'-V} &=& 4 \pi^2 t^4 \rho^2
f_{\rm loc}^2 ( 1- P^2)\\
\frac{I_{CAR}}{V'-V} &=& 4 \pi^2 t^4 \rho^2
\LL f^2_{\alpha,\beta} \RR (1+P^2)
\label{eq:ICAR-sym-SS}
,
\end{eqnarray}
where we used the notation $\tilde{\rho}=\rho/2$
for the spin-up or spin-down density of state in the
ferromagnetic electrodes. Eqs.~(\ref{eq:IAR-sym-SS})
and~(\ref{eq:ICAR-sym-SS}) in the limit $P=1$ are in
agreement with Eq.~(\ref{eq:Itot-half-SS}) in the
limit of a symmetric contact.
In the symmetric case the conductance is thus 
equal 
to the conductance associated to a single superconductor
divided by two, in agreement with Eq.~(\ref{eq:series-SS}).


\section{Josephson effect in a two-channel ballistic S/DW/S junction}
\label{app:small-dis}
In this appendix we describe the Josephson
effect with a ballistic propagation in the ferromagnetic electrodes.
In the limit of a long junction $R \gg a_0$ the matrix
$\CHI-\CHK^{A,R}$ given by (\ref{eq:K4x4})
is block-diagonal because the
Andreev bound states do not couple the two superconductors.
There exist two bound states associated
to the interfaces $(\alpha,a)$ and $(\beta,b)$ and
two bound states associated to the interfaces
$(\alpha',a')$ and $(\beta',b')$.
The secular equation for the bound states living at
the interfaces
$(\alpha,a)$ and $(\beta,b)$ takes the form
\be
\label{eq:secular}
1+i\pi^2 \rho_F \rho_N (t_\alpha^2+t_\beta^2)
\frac{\omega_0}{\sqrt{\Delta^2-\omega_0^2}}
+ (\pi^2 \rho_F \rho_N)^2 t_\alpha^2 t_\beta^2
\frac{z^2 \Delta^2 - \omega_0^2}{\Delta^2-\omega_0^2} =0
,
\ee
where $\rho_N$ and $\rho_F$ are the density of states in the
superconductor and in the half-ferromagnetic electrodes, and
where we used the notation
$z=\sin[k_F R]/(k_F R)$.
In the case $t=t_\alpha=t_\beta$ and in the tunnel limit
$\pi t^2 \rho_N \rho_F  \ll 1$ the solution of (\ref{eq:secular})
takes the form
\be
\omega_0^2 = \Delta^2 \left[ 1 +
(\pi^2 t^2 \rho_N \rho_F)^2 (1 \pm z)^2 \right]
.
\ee
The supercurrent is easily deduced from (\ref{eq:super-SDWS}):
\begin{eqnarray}
I_S&=& \frac{16 \pi^9}{ (a_0 k_F^\uparrow)^2
(a_0 k_F)^2} 
\frac{e}{h} \Delta t^4
\rho_N^2 \rho_F^2 (1+P)^2
\frac{a_0}{2\pi R_{\alpha,\beta}}
\frac{a_0}{2\pi R_{\alpha',\beta'}}
\frac{a_0}{2\pi R_{a,a'}}
\frac{a_0}{2\pi R_{b,b'}}
\sin{[k_F R_{\alpha,\beta}]}
\sin{[k_F R_{\alpha',\beta'}]}\\
&& \exp{\left\{-\left(
\frac{R_{a,a'}+R_{b,b'}}{l_\phi}\right)\right\}}
f(z,z')
\cos{\alpha} \sin{\varphi}
,
\end{eqnarray}
where $\alpha$ is defined by
\be
\alpha = k_F^\uparrow(R_{a,a'}-R_{b,b'})
+\frac{\Delta}{v_F^\uparrow} (R_{a,a'}+R_{b,b'})
,
\ee
and where $f(z,z')$ is a geometrical prefactor of
order unity.

\end{document}